\documentstyle[prl,aps,epsf,twocolumn,floats]{revtex}
\renewcommand{\refname}{}
\newcommand{\biblabel}[1]{[#1]} %
\renewcommand{\references}{%
\ifpreprintsty \vspace*{-0.1 truein}
\hbox to\hsize{\hss\large \refname\hss}%
\else
\vskip3pt
\hrule width\hsize\relax
\vskip -0.2in
\fi
\list{\biblabel{\arabic{enumiv}}}%
{\labelwidth\WidestRefLabelThusFar  \labelsep4pt %
\leftmargin\labelwidth %
\advance\leftmargin\labelsep %
\ifdim\baselinestretch pt>1 pt %
\parsep  4pt\relax %
\else %
\parsep  0pt\relax %
\fi
\itemsep\parsep %
\usecounter{enumiv}%
\def\theenumiv{\arabic{enumiv}}%
}%
\let\newblock\relax %
\sloppy\clubpenalty4000\widowpenalty4000
\sfcode`\.=1000\relax
\ifpreprintsty\else\small\fi
}

\begin{document}

\draft

\title{
Field dependent equilibrium energetic distribution of
localized charge carriers in disordered semiconductors
at low temperatures.
}
\author{D.V.~Nikolaenkov,~V.I.~Arkhipov$^*$ and V.R.~Nikitenko}

\address{
Moscow Institute of Physics and Engineering,
Kashirskoye shosse 31, Moscow 115409, Russia.
\\
$^{*}$ IMEC, Kapeldreef 75, B-3001 Heverlee-Leuven, Belgium}

 \maketitle

\begin{abstract}
The hopping transport of charge carries in 3-dimensional
disordered semiconductors is analyzed in this work. It is assumed
that the density of localized states is a decreasing function of
energy and contribution of thermoactivated jumps into the
transport is negligible. The possibility of equilibrium energetic
distribution for this case is shown. This distribution is
characterized by field--dependent effective temperature.
\end{abstract}

\centerline{}

Processes of carrier transport in amorphous semiconductors with
conduction via delocalized states have been successfully described
in terms of the multiple-trapping model [1-3]. This model implies
that two carrier fractions exist in the material, occupying at any
given instant of time delocalized and localized states. However,
in the materials with all carriers in localized states, e.g., in
most polymers, charge transport is due to carrier hopping. Many
interesting and important features of the hopping transport in
disordered semiconductors have been observed experimentally in
fairly high electric fields [4-6]. It has been shown both
analytically [7-10] and by Monte Carlo simulations [11, 12] that
the energy distribution of localized carriers and the carrier
mobility in a disordered semiconductor can be described both by
the ordinary temperature $T$ and by some effective temperature
$T_F(F)$ dependent on the magnitude of the applied external
electric field $F$. Analytical calculations [7, 9] of this
effective temperature yield
 $$ T_F= {eF\over 2\gamma k},\eqno(1)$$
  while numerical simulation yields $T_F=0.67\cdot eF/\gamma k$
  (see [11]) or
   $T_F=(0.69\pm 0.03)\cdot eF/\gamma k$
   (see [12]), where $e$ is the elementary charge, $1/\gamma$ the
localization radius, and $k$ the Boltzmann constant. The
equilibrium energy distribution of injected carriers in an
amorphous semiconductor subjected to an external electric field at
low temperatures, when the contribution of thermally activated
hops to the transport process is negligible, has been found [9]
for the 1D case, which, in fact, corresponds to high electric
fields. In this paper, we demonstrate that the equilibrium energy
distribution of localized carriers is also possible in the 3D case
at lo w temperatures in an external electric field. We show that
this distribution can be described by the Boltzmann exponential
with an effective temperature given by (1).

As an equation describing the kinetics of the hopping transport,
we use the well--known balance equation
 $${\partial f_i \over
\partial t} = \sum\limits_{j\not=j} \nu_{ji} f_j - f_i
\sum\limits_{j\not=j} \nu_{ij}, \eqno(2)$$
 written for the case of
low occupancy of localized states, $f_i\ll 1$. Here, $t$ is the
time, $f_i$ the mean occupation number of the $i$-th state and
$\nu_{ij}$ the probability of transition from the state $i$ to the
state $j$ . Let us assume that in the quasiequilibrium transport
regime, when the density of states is time--independent, a great
number of carriers can a void unlikely jumps that take a very long
time $t_R\gg \hbar /\Delta E$, i.e., jumps that do not contribute
significantly to the transport process [13]. Here, $\hbar$ is
Planck's constant and $\Delta E$ the characteristic energy change
in carrier transition from one localized state to another. This
assumption makes it possible to take advantage of the concept of
the distribution function $f\left({\bf r},E,t\right)$ averaged
over continuous variables: energy $E$ (''deeper'' states have
higher energies $E$) and radius vector ${\bf r}$. Let us use the
Miller--Abrahams expression for the transition rate
 $$\nu(|\Delta{\bf r}|,E)=\nu_0 \exp\left(-2\gamma |\Delta{\bf
r}|-H({\cal E}){{\cal E}\over kT}\right), \eqno(3)$$
 where $ {\cal
E} \equiv E-e{\bf F}\Delta{\bf r}-E' $, $H({\cal E})$ is the
Heaviside function, $\nu_0$ the attempt-to-jump frequency, and
$\Delta{\bf r}={\bf r}-{\bf r'}$. Then, we use equation derive the
following kinetic equation for the distribution function $f$,
 $$
{\partial f\left({\bf r},E,t\right)\over \partial t} = \nu_0 \int
\,d{\bf r'}\exp\left(-2\gamma |\Delta{\bf r}|\right) g(E) \times
 $$
 $$ \Biggl[ \int^{\infty}_{E-e{\bf F}\Delta{\bf
r}}\,dE'\exp\biggl( {E-e{\bf F} \Delta{\bf r}-E'\over kT}\biggr)
f\left({\bf r'},E',t\right)$$
 $$ + \int_{-\infty}^{E-e{\bf
F}\Delta{\bf r}}\,dE' f\left({\bf r},E',t\right)
 \Biggr] -$$
 $$ \nu_0\int \,d{\bf r'}\exp\left(-2\gamma |\Delta{\bf r}
|\right) f\left({\bf r} ,E,t\right) \times$$
 $$
\Biggl[\int_{-\infty}^{E-e{\bf F}\Delta{\bf r}}\,dE'\exp\biggl(
{-\left[E-e{\bf F} \Delta{\bf r}-E'\right]\over kT}\biggr) g(E')
 $$
$$ +  \int^{\infty}_{E-e{\bf F}\Delta{\bf r}}\,dE' g(E')\Biggr],
\eqno(4) $$
 by the condition
  $$ \int\,d{\bf r}
\int^{\infty}_{-\infty}f\left({\bf r},E,t\right)dE = 1, \eqno(5)
 $$
 where $g(E)$ is the energy distribution of localized states
(DOS) normalized by the condition $$ \int_{-\infty}^\infty\,g(E)dE
= 1. \eqno(6) $$ The zero energy $E=0$ corresponds to the maximum
of the function $g(E)$. Equation (4) is written on the assumption
that the position and energy of localized states are uncorrelated,
which corresponds to the case of completely disordered materials.

At low temperatures, when the contribution of thermally activated
hops can be neglected subject to the condition of hypothetical
equilibrium, kinetic equation (4) for the equilibrium distribution
function averaged over the space coordinates
 $$
f_{eq}\left(E\right) = \int f\left({\bf r},E,\infty\right)\,d{\bf
r} \eqno(7) $$
 takes the form
 $$ 0 =  \nu_0\int \,d{\bf
r'}\exp\left(-2\gamma |\Delta{\bf r}|\right)
\int_{-\infty}^{E-e{\bf F}\Delta{\bf r}}\,dE'
f_{eq}\left(E'\right) g(E) \ - $$
 $$\nu_0\int \,d{\bf r'}
 \exp\left(-2\gamma |\Delta{\bf r}|\right) \int^{\infty}_{E-e{\bf F}
\Delta{\bf r}}\,dE' g(E') f_{eq}\left(E\right). \eqno(8) $$
 We use
equation (8) and formulas (5) and (6) to derive the following
integral equation:
 $$ f_{eq}\left(\varepsilon\right) =
g(\varepsilon) { A\left(\varepsilon,\varepsilon'\right)
f_{eq}\left(\varepsilon'\right) \over
4-A\left(\varepsilon,\varepsilon'\right)g(\varepsilon') }.\eqno(9)
$$
 Here, we introduced the integrated operator
  $$A\left(\varepsilon,\varepsilon'\right)= \int_{-\infty}^{\infty}
\,d\varepsilon' \biggl[4 H(\varepsilon-\varepsilon') - $$
  $$
sgn(\varepsilon-\varepsilon')
\bigl(2+|\varepsilon-\varepsilon'|\bigr)e^{-|\varepsilon-\varepsilon'|}
 \biggr], \eqno(10)$$
 and the dimensionless variables $\varepsilon =
E/kT_F$ and $\varepsilon' = E'/kT_F$, where $T_F$ is the effective
temperature defined by (1). The appearance of this effective
temperature is associated with the fact that, during transport,
the electric field transfers carriers into states with a higher
energy, similarly to the thermodynamic effects described by the
ordinary temperature $T$.

This equation can be solved by the iteration method; as the first
approximation, we use the density of localized states $$
f^{(0)}_{eq}(E) = g(E). \eqno(11)  $$ Substituting (11) into (9),
we obtain
 $$
A\left(\varepsilon,\varepsilon'\right)g(\varepsilon')\sim \cases{
4-\varepsilon e^{-\varepsilon},&$E\gg E_0+E_m\left(T_F\right)$;\cr
-\varepsilon e^\varepsilon,&$E\ll
-E_0-E_m\left(T_F\right)$.\cr}\eqno(12)$$
 Here, the quantity
$E_m(T_F)$ depends on the particular form of the DOS function. In
particular, in the case of a Gaussian distribution
 $$g(E)=g_0\exp\left(-[E/E_0]^2\right),$$
  we have
$E_m\left(T_F\right) = E_0^2/2T_F,$ while in the case of an
exponential distribution
 $$ g(E)=g_0\exp\left(-|E/E_0|\right) $$
we obtain $E_m\left(T_F\right) = E_0^2/(T_F-E_0).$

Hence,
 $$ f^{(1)}_{eq}\left(E\right) \sim |E|^{-sgn(E)}g(E)
\exp\left({E\over kT_F}\right), $$
 $$ |E| \gg E_0 +
E_m(T_F).\eqno(13) $$
 Thus, to a first approximation, the
equilibrium energy distribution is described by the Boltzmann
function. To calculate the distribution function $f_{eq}(E)$ more
precisely, we substitute, in accordance with (13), the function
$g(E)exp(E/kT_F)$ into the right--hand side of equation (9).The
resulting solution $f^{(2)}_{eq}(E)$
 has the same asymptotic form as the function $f^{(1)}_{eq}(E)$, since the
asymptotic form (12) of the function
$A\left(\varepsilon,\varepsilon'\right)g(\varepsilon')$ appearing
in equation (9) is not changed. Thus, if the DOS function $g(E)$
is sufficiently ''shallow'', so that the equilibrium can be
established,
 $$E \, g(E)\exp\left({E\over kT_F}\right)\to 0, \ \ \ \
\ \ |E|\to\infty, \eqno(14)$$
 then the asymptotic form of the
occupancy of localized states $\varphi(E)=f(E)/g(E)$ is the
exponential $\exp(E/kT_F)$. The quantity $E_m(T_F)$ appearing in
equations (12) and (13) is the characteristic energy of the peak
in the distribution function $f_{eq}(E)$.

Thus, we showed the following. First, in the 3D case, as also in
the 1D one [9], the spatially uniform equilibrium energy
distribution of localized charge carriers may occur even in the
limiting case $T\to 0$, when the contribution from thermally
activated hops to the kinetic equation (4) can be completely
neglected. Second, this distribution can be approximated by the
Boltzmann distribution function with the effective temperature (1)
instead of the ordinary temperature. It should be noted that this
effective temperature coincides with previous analytical
estimations [9, 10].

\end{document}